\documentclass[11pt,a4paper]{article}
\pdfoutput=1

\usepackage[]{Packages}

\bibliography{GaugeBetheRef}

\usepackage{Definitions}

\allowdisplaybreaks[1]

\newcommand*{\dual}[1]{\tensor[^*]{#1}{}}

\preprint{\texttt{CERN-PH-TH/2013-073}}

\newcommand{\OfficialTitle}{Omega--Deformed Seiberg--Witten Effective Action from the M5--brane}

\title{\vspace{2cm}
  {\huge   \textbf{\dosserif\OfficialTitle}}
}

\hypersetup{pdfauthor={and Domenico Orlando and Susanne Reffert},pdftitle={\OfficialTitle}}

\author{%
  \begin{minipage}{.8\linewidth}
    \vspace{1cm}
    \begin{center}
      {\small \textbf{Neil Lambert}\footnote{On leave from King’s College London.}{}, \textbf{Domenico Orlando} and \textbf{Susanne Reffert}}
    \end{center}
    \vspace{1cm}
    \begin{minipage}{\linewidth}\centering
      {\itshape \footnotesize
        Theory Group, Physics Department, \\ Organisation européenne pour la recherche nucléaire (CERN) \\ CH-1211 Geneva 23, Switzerland
      }
    \end{minipage}
  \end{minipage}
}

\date{}

\begin{document}

\setstretch{1.1}

\numberwithin{equation}{section}

\begin{titlepage}

  \maketitle

  \thispagestyle{empty}

  \vfill
  \abstract
{We obtain the leading order corrections to the effective action of an \M5--brane wrapping a Riemann surface in the eleven-dimensional supergravity $\Omega$--background. The result can be identified with the  first order  $\epsilon$--deformation of the Seiberg--Witten effective action of pure $SU(2)$ gauge theory. We also comment on the second order corrections and the generalization to arbitrary gauge group and matter content.}
\vfill

\end{titlepage}

\section{Introduction}

Ever since the classic result of \ac{sw}~\cite{Seiberg:1994rs}, $\mathcal{N}=2$ gauge theories have occupied a prominent place in theoretical physics. The resulting low energy \ac{sw} effective action  is given in terms of a Riemann surface, the \ac{sw} curve, which encodes all the perturbative and non-pertubative quantum effects of the gauge theory. While all the perturbative corrections had been known since~\cite{Gates:1981yc,Howe:1983wj,Seiberg:1988ur}, this solution gave a prediction for an infinite number of non-perturbative instanton corrections, the first few terms of which could be checked by explicit computation~\cite{Dorey:1996hu,Dorey:1996bf}.

Not long afterwards, M--theory was developed as an eleven-dimensional non-per\-tur\-ba\-tive completion of String Theory. In a striking paper  Witten showed how the \ac{sw} curve could be naturally obtained from the geometry of intersecting  \NS5 and \D4--branes lifted to M--theory where they become a single \M5--brane~\cite{Witten:1997sc}. Moreover the complete quantum \ac{sw} effective action for $\mathcal{N} = 2$ supersymmetric $SU(N)$ Yang--Mills theory was obtained in~\cite{Howe:1997eu} from the classical dynamics of the \M5--brane.

An alternative method to compute the \ac{sw} solution from first principles came with Nekrasov's seminal paper using the $\Omega$--background~\cite{Nekrasov:2002qd}. This background deforms the gauge theory and allows for localization techniques to be used to compute all the instanton corrections and also reconstruct the curve and its associated quantities~\cite{Nekrasov:2003rj}. Since then the $\Omega$--background has received a lot of interest, most recently in the context of the correspondence by Alday, Gaiotto and Tachikawa~\cite{Alday:2009aq} and work related to it.  

The so-called \emph{fluxtrap} background~\cite{Hellerman:2011mv,Reffert:2011dp,Hellerman:2012rd} provides a string-theoretical construction of the Euclidean  $\Omega$--background determined by a  two-form $\omega = \di U$.\footnote{See~\cite{Fucito:2011pn,Antoniadis:2013bja} for alternative realizations.} In particular the bosonic  Abelian worldvolume  action for   \D4--branes suspended between  \NS5--branes in this background was given in \cite{Hellerman:2012zf}. The generalization to non-Abelian fields is given by ($\mu,\nu=0,1,2,3$) 
\begin{multline}
  \label{action}
  \mathscr{L}_{\D4} = \frac{1}{g^2_4} \Tr \Big[\frac{1}{4} \mathbf{F}_{\mu\nu} \mathbf{F}_{\mu\nu} +  \frac{1}{2}(\mathbf{D}_\mu \, \boldsymbol{\varphi} + \frac{1}{2} \mathbf{F}_{\mu\lambda}\hat U^\lambda)(\mathbf{D}_\mu \, \bar{\boldsymbol{\varphi}}  + \frac{1}{2}\mathbf{F}_{\mu\rho}\hat U^\rho) \\ 
- \frac{1}{4}[\boldsymbol{\varphi}, \bar{\boldsymbol{\varphi}}]^2+ \frac{1}{8}(\hat U^\mu \mathbf{D}_\mu (\boldsymbol{\varphi}-\bar{\boldsymbol{\varphi}}))^2\Big] \ ,
\end{multline}
where a hat denotes the pull-back to the brane and a bold-face indicates a non-Abelian field.
 The \emph{fluxtrap} can be lifted to M--theory~\cite{Hellerman:2012zf}. At order $\epsilon$ it is given by  ($M,N=0,1,2,...,10$)
\begin{subequations}
  \label{eq:first-order-Mbulk}
  \begin{align}
     g_{MN} &= \delta_{MN} + \mathcal{O}(\epsilon^2) \, ,\\
    G_4 &= \left( \di z + \di \bar z  \right) \wedge \left( \di s + \di \bar s \right) \wedge \omega\, ,
  \end{align}
\end{subequations}
where \( s = x^6 + \im x^{10} \), \( z = x^8 + \im x^9 \), and
\begin{equation}
  \omega = \epsilon_1 \di x^0\wedge \di x^1 + \epsilon_2 \di x^2\wedge \di x^3 + \epsilon_3 \di x^4 \wedge \di x^5 \ .
\end{equation}
The background has 8 Killing spinors if \( \epsilon_1 + \epsilon_2 + \epsilon_3 = 0 \), and 16 Killing spinors in the special case \( \epsilon_1 = - \epsilon_2 \) and \( \epsilon_3 = 0 \).\footnote{The $\epsilon_3$ component, although generically non-vanishing, will not play a role in this paper as the \M5--brane will be held fixed in the $x^4,x^5$ plane.}

In this paper we will derive the corrections to first order in $\epsilon$ to the $\Omega$--deformed \ac{sw} action. We do this by employing the M--theory lift of the fluxtrap background. As we will see, the classical M--theory calculation has the invaluable benefit of giving a quantum result in gauge theory since in this case, the result is independent of the effective coupling in the gauge theory. We embed the \M5--brane in the $\Omega$-background and study the most supersymmetric configuration which to first order in $\epsilon$ is still of the form $\setR_4 \times \Sigma$ with an additional self-dual three-form. %
This is the ground state of a six-dimensional theory on top  of which we have fluctuations fulfilling some assumptions detailed in the following. %
These fluctuations obey \emph{scalar} and \emph{vector equations of motion} that arise from the six-dimensional theory, where the scalar equation encodes the fact that the M5--brane is a (generalized) \emph{minimal surface} and the vector equation posits that the self-dual three-form on the brane is the (generalized) \emph{pullback of the three-form field in the bulk}. To arrive at the four-dimensional gauge theory, we must integrate these equations over the Riemann surface $\Sigma$ %
using an appropriate measure.
The integration results in one vector equation and two scalar equations in four dimensions, which are the Euler--Lagrange equations for a four-dimensional action, which in the case $\epsilon=0$ reproduces the undeformed \ac{sw} action.
We explicitly treat the case of \( SU(2) \) without matter, however there is a natural generalization of our result to any gauge group and matter content.

\bigskip
The plan of this paper is as follows. In Section~\ref{sec:m5} we describe the embedding of the \M5--brane, the six-dimensional equations of motion and their reduction to four-dimensions. We also give an action that captures these equations of motion. This action can be extrapolated to second order in $\epsilon$ and  generalized to arbitrary gauge group and matter content. In Section~\ref{sec:conclusions} we give our conclusions. We also provide an appendix that gives some technical steps in the evaluation of various non-holomorphic integrals over the Riemann surface that arise.

\section{M5--brane dynamics in the $\Omega$--fluxtrap}\label{sec:m5}

\paragraph{The homogeneous embedding of the \M5--brane.}

 Due to the fundamentally Euclidean nature of the fluxtrap background, we will be discussing the Euclidean version of \ac{sw}-theory. For this reason, the self-duality condition for the three-form \( h_3 \) on the \M5--brane turns into
\begin{equation}
    \im*_6 h_3 =  h_3 \,,
\end{equation}
which we will refer to as \emph{self-duality}.

The embedding of the \M5--brane in the fluxtrap background at order \( \epsilon \) has already been discussed in~\cite{Hellerman:2012zf}, where it was found that the brane wraps a Riemann surface. Let us recall here the argument.
As discussed in~\cite{Witten:1997sc}, the M--theory lift of a \NS5--\D4 system (extended respectively in \( x^0, \dots, x^3, x^8, x^9 \) and \( x^0, \dots, x^3, x^6 \)) is a single \M5--brane extended in \( x^0, \dots, x^3 \) and wrapping a two-cycle in \( x^6, x^8, x^9, x^{10} \). We use static gauge and assume that the \M5--brane has coordinates $x^\mu$, $\mu=0,1,2,3$ and $z=x^8+\im x^9$. We also assume that the only non-vanishing  scalar field is $s = x^6+\im x^{10}$. The precise form of the embedding is found if we require this brane to preserve the same supersymmetries of the original \tIIA system. Given the Killing spinors \( \eta_0 \) of the bulk, the \M5--brane preserves those satisfying~\cite{Howe:1996yn,Howe:1997fb} ($m,n=0,1,2,...,5$)
\begin{align}
  \proj{\M5}_- \eta_0 &= \tfrac{1}{2} \left( 1 - \Gamma_{\M5}  \right)  \eta_0 = 0 \ , &
  \Gamma_{\M5} &= -\frac{\epsilon^{m_1 \dots m_6} \hat \Gamma_{m_1 \dots m_6} }{6! \sqrt{\hat g}}
  \left( 1 - \tfrac{1}{3} \hat \Gamma^{n_1 n_2 n_3} h_{n_1 n_2 n_3} \right)  \ ,
\end{align}
where \( \hat \Gamma \) and \( \hat g \) are the gamma matrices and the metric, pulled back to  the brane. Here \( h_3 \) is the self-dual three-form on the \M5--brane worldvolume which satisfies
\begin{equation}
  \di H_3 = - \tfrac{1}{4}  \hat G_4 \,,
\end{equation}
where $H_3 = h_3 + {\mathcal O}(h_3^3)$.

For \( \epsilon = 0 \) we have $h_3=0$ and the \M5--brane is described by a Riemann surface \( \bdel s = 0 \)~\cite{Witten:1997sc}.
Let us now consider the first order effect that arises when turning on $\epsilon$. To this order we may simply take $H_3=h_3$ but in principle $s$ may pick up a non-holomorphic piece. However at ${\mathcal O}(\epsilon)$
the pullback  only  depends holomorphically on \( s (z) \) since \( \hat \omega \) is by itself of order \( \epsilon \): 
\begin{equation}
  \hat G_4  = - \left( \del s  - \bdel \bar s  \right) \di z \wedge \di \bar z \wedge \hat\omega+{\mathcal O}(\epsilon^2)\, .
\end{equation}
Therefore we can take
\begin{equation}
  h_3 = \tfrac{1}{4} \left( \bar s  - \bar z \del s  + f(z)  \right) \di z \wedge \hat\omega^- + \tfrac{1}{4}  \left(  s  -  z \bdel \bar s  + \bar f(\bar z) \right) \di \bar z \wedge \hat\omega^+ \, ,
\end{equation}
where $f$ is an arbitrary holomorphic function and we have decomposed the two-form \( \hat\omega \) as
\begin{equation}
\begin{aligned}
  \hat\omega &= \frac{\epsilon_1 + \epsilon_2}{2} \left( \di x^0 \wedge \di x^1 + \di x^2 \wedge \di x^3 \right) + \frac{\epsilon_1 - \epsilon_2}{2}  \left( \di x^0 \wedge \di x^1 - \di x^2 \wedge \di x^3 \right) \\
  &= \hat\omega^+ + \hat\omega^- \, .
\end{aligned}
\end{equation}
These are all the ingredients needed to write the supersymmetry condition,
\begin{equation}
  \proj{\M5}_- \eta = \proj{\M5}_- \proj{\NS5}_+ \proj{\D4}_+ \eta_0 = 0\,,
\end{equation}
where the projectors \( \proj{\NS5} \) and \( \proj{\D4} \) refer to the \M5--branes resulting from the lift of the \NS5--brane and \D4--brane introduced above such that \( \eta = \proj{\NS5}_+ \proj{\D4}_+ \eta_0 \) are the Killing spinors preserved by the branes. Since the two \M5--brane projectors commute, the full configuration preserves two supercharges in the generic case and four if \( \epsilon_1 = - \epsilon_2 \).  An explicit calculation shows that the condition is satisfied at ${\mathcal O}(\epsilon)$ if
\begin{equation}
  \begin{cases}
    \bdel s   = 0\ , \\
    f(z) = 0 \ ,
  \end{cases}
\end{equation}
which completely fix the embedding of the \M5--brane and the self-dual field \( h_3 \).

Thus even at order \( \mathcal{O}(\epsilon) \) the brane is embedded holomorphically in spacetime. For the simplest case corresponding to pure $SU(2)$ Yang--Mills, the precise form was found in~\cite{Witten:1997sc} and is determined implicitly by
\begin{align}
  t^2- 2B(z | u)t + \Lambda^4 &= 0 \ , & t &= \Lambda^2e^{- s/R}
 \ ,
\end{align}
where  \( B(z | u) = \Lambda^4z^2 - u\), $\Lambda$ is a mass scale and $R$ the radius of the $x^{10}$-direction. This embedding defines a Riemann surface $\Sigma$ with modulus \( u \),
\begin{equation}
  \Sigma = \set{(z,s)| s = s(z|u)} .
\end{equation}
It is useful to observe that
\begin{equation}
 \frac{ \del s}{\del u} \di z= -\frac{1}{2\Lambda^4z} \frac{ \del s}{\del z} \di z =   \frac{R \di z}{\sqrt{Q(z|u)}} =  R\lambda
\end{equation}
is the unique holomorphic one-form on $\Sigma$
where  \( Q(z|u) = B(z|u)^2 - \Lambda^4 \).  For most of this paper we will simply set $R=\Lambda=1$. They are in principle needed on dimensional grounds, since both $s$ and $z$ have dimensions of length whereas the modulus $u$ is usually taken to have mass-dimension two. We will briefly reinstate them in the conclusions by simply rescaling $z$ and $s$, when discussing the quantum nature of our result.

\paragraph{Equations of motion in 6d.}

Having found the embedding of the \M5--brane we want to describe the low-energy dynamics of the fluctuations around the equilibrium. In fact, since we are interested in the effective four-dimensional theory living on \( x^0, \dots, x^3 \) which results from integrating the \M5 equations of motion over the Riemann surface \( \Sigma \), we will assume that:
\begin{enumerate}
\item the geometry of the five-brane is still a fibration of a Riemann surface over \( \setR^4 \);
\item for each point in \( \setR^4 \) we have the same Riemann surface as above, but with a different value of the modulus \( u \).
\end{enumerate}
In other words, the modulus \( u \) of \( \Sigma \) is a function of the worldvolume coordinates and the embedding is still formally defined by the same equation, but now \( s = s ( z | u(x^\mu)) \) so that the \( x^\mu \)--dependence is entirely captured by
\begin{equation}
  \label{eq:xmu-dependence}
  \del_\mu s (z | u (x^\mu)) = \del_\mu u \frac{\del s}{\del u} \,.
\end{equation}
For ease of notation we will drop in the following the explicit dependence of \( s \) on \( u(x^\mu) \) and write directly \( s = s (z, x^\mu) \). Much of our discussion follows the undeformed case considered in detail in~\cite{Howe:1997eu,Lambert:1997dm,Lambert:1997he}.

The dynamics can be obtained by evaluating the \M5--brane equations of motion. Here we will only focus on the bosonic fields. Covariant equations of motion for the \M5--brane were obtained in~\cite{Howe:1996yn,Howe:1997fb}. In general these are rather complicated equations, particularly with regard to the three-form. However in this paper we only wish to work to linear order in $\epsilon$ and quadratic order in spatial derivatives $\del_\mu$. In particular we can take $H_3=h_3$ and the equations of motion reduce to\footnote{Note that we have chosen the opposite sign to the \emph{rhs} of the scalar equation as compared to what is given in~\cite{Howe:1997fb}. This corresponds to a choice of brane or anti-brane.}
\begin{align}
\label{eq:general-scalar-equation}
 \left( \hat g^{mn} - 16 h^{mpq}h^n{}_{pq} \right) \nabla_m \nabla_n X^{I} &=  -\frac{2}{3 }  \hat G\indices{^{I}_{mnp}} h^{mnp} \,, \\
  \di h_3 &= - \tfrac{1}{4} \hat G_4 \,,
\end{align}
where $I = 6,\dots,10$ and the geometrical quantities are defined with respect to  the pullback of the spacetime metric to the brane ${\hat g}_{mn}$.

As a first step we need to write the three-form field on the brane. In full generality, \( h_3 \) can be decomposed as
\begin{equation}
  h_3 = - \tfrac{1}{4} \left( \hat C_3 + \im *_6 \hat C_3 - \Phi \right)  \, ,
\end{equation}
where \( \hat C_3 \) is the pullback of the three-form in the bulk, and \( \Phi \) is a self-dual three-form that will encode the fluctuations of the four-dimensional gauge field.

Since we ultimately want to discuss the gauge theory living on the worldvolume coordinates \( x^0, \dots, x^3 \), we make the following self-dual ($\im*_6 \Phi =   \Phi$) ansatz for \( \Phi \):
\begin{equation}
  \begin{aligned}
    \Phi ={}&\frac{ \kappa}{2}\cF_{\mu\nu}\di x^\mu\wedge \di x^\nu\wedge\di z+ \frac{\bar \kappa}{2}\wt{\cF}_{\mu\nu}\di x^\mu\wedge \di x^\nu\wedge\di \bar z\\
    & +  \frac{1}{1+|\del s|^2}\frac{1}{3!}\epsilon_{\mu\nu\rho\sigma}\left(  \del^\tau s \bdel \bar s \, \kappa \cF_{\sigma\tau}-\del^\tau\bar s\del s \, \bar\kappa \wt{\cF}_{\sigma\tau} \right)\di x^\mu\wedge\di x^\nu\wedge\di x^\rho \,.
  \end{aligned}
\end{equation}
The two-form \( \cF \) is anti-self-dual in four dimensions, while \( \wt{\cF} \) is self-dual:
\begin{align}
 *_4 \cF &= - \cF \,, & *_4 \wt{\cF} &= \wt{\cF} \,.
\end{align}
Here $*_4$ is the flat space Hodge star and \( \kappa(z) \) is a holomorphic function given by~\cite{Lambert:1997dm}
\begin{equation}
  \kappa = \frac{\di s}{\di a} =
  \left( \frac{\di a}{\di u}  \right)^{-1} \lambda_z \, .
\end{equation}
Here $\lambda = \lambda_z dz$ is the holomorphic one-form on $\Sigma$ and \( a \) is the scalar field used in the Seiberg--Witten solution and related to $\lambda$ by
\begin{equation}
  \frac{\di a}{\di u} = \oint_A \lambda\ ,
\end{equation}
where $A$ is the a-cycle of $\Sigma$. In the following,  ${\cF}$ and $\wt{\cF}$ will be related to the four-dimensional gauge field strength, thus justifying our ansatz.

We also need to choose a gauge for the three-form potential \( C_3 \) in the bulk:
\begin{equation}
    C_3 = -\tfrac{1}{2} \left(\bar s\di v-\bar v\di s+s\di v-\bar v \di \bar s\right)\wedge\omega+\mathrm{c.c}.
\end{equation}
Its pullback on the Riemann surface $\set{ v = z, s = s(z, x^\mu)}$ is given by
\begin{equation}
  \hat C_3 = -\tfrac{1}{2} \left(\bar s\di z-\bar z\del s\di z-\bar z\del_\mu s\di x^\mu+s\di z-\bar z \bdel \bar s\di\bar z-\bar z\del_\mu\bar s\di x^\mu\right)\wedge\hat\omega + \text{c.c.} \, .
\end{equation}
We are only interested in terms up to second order in the spacetime derivatives \( \del_\mu \) and in particular we observe that \( \hat\omega \) is by itself of first order. It follows that the six-dimensional Hodge dual is given by
\begin{equation}
  \begin{aligned}
    \im *_6\hat C_3 ={}& \tfrac{1}{2} \left(\bar s\di z-\bar z\del s\di z+s\di z+\bar z\bdel s\di\bar z - s \di \bar z + z \bdel\bar s \di\bar z - \bar s \di\bar z - z \del s \di z \right) \wedge \dual{\hat \omega}\\
    &   +\frac{1}{2\cdot 3!}\left( 1 + \abs{\del s}^2 \right)\epsilon_{\mu\nu\lambda\rho}C^{\mu\nu\lambda} \di x^\rho\wedge\di z\wedge \di\bar z \\
    &  +\frac{1}{1+\abs{\del s}^2}\epsilon_{\mu\nu\rho\sigma}\left(\del^\tau s \bdel \bar s \hat C_{\sigma \tau z} -  \del^\tau\bar s\del
      s   \hat C_{\sigma\tau\bar z} \right) \di x^\mu \wedge \di x^\nu \wedge \di x^\rho \ ,
  \end{aligned}
\end{equation}
where \( \dual{\hat\omega} = *_4 \hat\omega = \hat\omega^+ - \hat\omega^- \).

\paragraph{The vector equation.}

Consider now the vector equation \( \di h_3 = - \frac{1}{4} \hat H_4  \). Given our expression for \( h_3 \), the equation becomes
\begin{equation}
  \di \Phi = \im \di *_6 \hat C_3  \ ,
\end{equation}
where we see explicitly the role of the bulk three-form as source for the gauge field on the brane.
At this point it is useful to quickly discuss  the issue of gauge covariance of the three-form equation. The bulk three-form is defined up to the differential of a two-form \( C_3 \mapsto C'_3 + \di B_2 \). Under this shift the vector equation becomes
\begin{equation}
  \di \Phi =  \im  \di *_6 \hat C_3' + \im \di *_6 \di \hat B_2  \ ,
\end{equation}
which can be compensated for by an analogous shift in the fluctuations:
\begin{equation}
  \Phi \mapsto \Phi' + \di \hat B_2 + \im *_6 \di \hat B_2 \,.
\end{equation}

\bigskip

Let us go back to our ansatz. The tensor \( \Phi \) does not contribute to the \( \mu\nu z \bar z \) component:
\begin{equation}
   \left.\di \Phi \right|_{\mu\nu z\bar z} \equiv 0
\end{equation}
so we only need to verify that
\begin{equation}
  \left. \di *_6 \hat C \right|_{\mu\nu z\bar z} = 0 \, ,
\end{equation}
which is satisfied up to terms of order $\mathcal{O} (\del_\mu)^3 $, taking into account the fact that $\hat\omega$ is by itself of order $\mathcal{O}(\del_\mu) $. Similarly, also the \( \mu\nu \lambda\rho\) component of the equation of motion is of higher order.

It is convenient to take the six-dimensional dual of the remaining terms and decompose them in coordinates:
\begin{equation}
  *_6\di (\Phi - \im *_6 \hat C_3) =  \tfrac{1}{2}E_{\mu z} \di x^\mu \wedge \di z +   \tfrac{1}{2}E_{\mu \bar z} \di x^\mu \wedge \di \bar z = 0 \, ,
\end{equation}
where explicitly
\begin{subequations}
  \begin{align}
    E_{\mu z} &= \del_\mu ( \kappa \cF_{\mu \nu }- \hat C_{\mu \nu z}) + \del \left[ \frac{\bdel \bar s \del_\nu s}{1 + \abs{\del s}^2} ( \kappa \cF_{\mu\nu} - \hat C_{\mu\nu z}) \right] - \del \left[ \frac{\del s \del_\nu \bar s}{1 + \abs{\del s}^2} ( \bar \kappa \wt{\cF}_{\mu\nu} - \hat C_{\mu\nu \bar z} )\right] \, , \\
    E_{\mu \bar z} &= \del_\mu ( \bar \kappa \wt \cF_{\mu \nu } - \hat C_{\mu \nu \bar z}) + \bdel \left[ \frac{\del s \del_\nu \bar s}{1 + \abs{\del s}^2} ( \bar \kappa \wt{\cF}_{\mu\nu} - \hat C_{\mu\nu \bar z}) \right] - \bdel \left[ \frac{\bdel \bar s \del_\nu s}{1 + \abs{\del s}^2} ( \kappa {\cF}_{\mu\nu} - \hat C_{\mu\nu z} )\right] .
  \end{align}
\end{subequations}
Note that because of the epsilon tensors in the definition of \( E_{\mu z} \), the equations only depend on \( \hat\omega \) and not on \( \dual{\hat\omega} \).

To obtain the equations of motion of the vector zero-modes in four dimensions we need to reduce these equations on the Riemann surface. In order for the integral to be well-defined everywhere on \( \Sigma \) we have only two possible choices for the integrand, depending on the (unique) one-form \( \lambda \) or its complex conjugate: 
\begin{subequations}
  \label{eq:integral-vector-equation}
  \begin{align}
    \int_\Sigma *_6  \di (\Phi - \im \di * \hat C_3) \wedge \bar \lambda &= \di x^\mu \wedge \int_\Sigma E_{\mu z} \di z \wedge \bar \lambda = 0 \, ,\\
    \int_\Sigma *_6 \di (\Phi - \im \di * \hat C_3) \wedge \lambda &=
    \di x^\mu \wedge \int_\Sigma E_{\mu \bar z} \di \bar z \wedge \lambda = 0
    \, .
  \end{align}
\end{subequations}
The explicit integration is relatively straightforward using the techniques explained in Appendix~\ref{sec:line-integrals}. The only non-vanishing integrals have been already evaluated in~\cite{Howe:1997eu,Lambert:1997he}:
\begin{align}
  I_0 &= \int_\Sigma \lambda \wedge \bar \lambda = \frac{\di a }{\di u}
  \left( \tau - \bar \tau \right) \frac{\di \bar a}{\di \bar u} \, ,\\
  K &= \int_\Sigma \bdel
  \left[ \frac{\lambda_z \bdel \bar s}{1 + \abs{\del s}^2} \right] \di \bar z \wedge \lambda = -
  \left( \frac{\di a }{\di u} \right)^2 \frac{\di \tau}{\di u} \, ,
\end{align}
where one uses the following definitions:
\begin{align}
  a &= \oint_A \lambda_{SW} \, , & a_D &= \oint_B \lambda_{SW} \, , & \tau &= \frac{\di a_D}{\di a} \, , & \lambda &= \frac{\del \lambda_{SW}}{\del u} \, ,
\end{align}
along with the Riemann bi-linear identity
\begin{equation}
\int \lambda\wedge \bar \lambda = \oint_B\lambda\oint_A\bar\lambda - \oint_A\lambda\oint_B\bar\lambda\;.
\end{equation}
The two integrals in Eq.~\eqref{eq:integral-vector-equation} become
\begin{subequations}
  \label{eq:integrated-vector-equations}
  \begin{align}
    \left( \tau - \bar \tau \right)
    \left( \del_\mu \cF_{\mu\nu} + \del_\mu a \, \hat\omega_{\mu\nu} \right) + \del_\mu \tau \cF_{\mu \nu} - \del_\mu \bar \tau \wt{\cF}_{\mu\nu} &= 0 \, ,\\
    \left( \tau - \bar \tau \right) \left( \del_\mu \wt{\cF}_{\mu\nu}
      + \del_\mu \bar a \, \hat\omega_{\mu\nu} \right) + \del_\mu \tau
    \cF_{\mu \nu} - \del_\mu \bar \tau \wt{\cF}_{\mu\nu} &= 0 \, .
  \end{align}
\end{subequations}
Taking the difference of the two equations we find 
\begin{equation}
  \label{eq:difference-vector-equations}
  \del_\mu ( \cF_{\mu\nu} - \wt{\cF}_{\mu\nu}) = - \del_\mu \left( a - \bar a \right) \hat\omega_{\mu\nu} \, ,
\end{equation}
which is solved by writing
\begin{equation}
  \begin{cases}
    \cF = \left( 1 - * \right) F - \left( a - \bar a \right) \hat\omega^- \,,\\
    \wt{\cF} = \left( 1 + * \right) F + \left( a - \bar a \right) \hat\omega^+\,,
  \end{cases}
\end{equation}
where \( F \) satisfies the standard Bianchi identity
\begin{equation}
  \di F = 0 \, ,
\end{equation}
and can be written as the differential of a one-form \( F = \di A \). In the following we will identify \( F \) with the four-dimensional gauge field and, in this sense, Eq.~\eqref{eq:difference-vector-equations} represents the correction to the Bianchi equations introduced by the \( \Omega \)--deformation. Substituting this condition into the first equation of~\eqref{eq:integrated-vector-equations}, we derive the final form of the four-dimensional vector equations:
\begin{multline}
  \label{eq:vector-equation}
  \left( \tau - \bar \tau \right) \left[ \del_\mu F_{\mu\nu} + \tfrac{1}{2} \del_\mu ( a + \bar a ) \hat\omega_{\mu\nu} + \tfrac{1}{2} \del_\mu (a - \bar a) \dual{\hat\omega}_{\mu\nu} \right] \\+ \del_\mu \left( \tau - \bar \tau\right) \left[ F_{\mu\nu} + \tfrac{1}{2} \left( a - \bar a \right) \dual{\hat\omega}_{\mu\nu} \right]  %
  - \del_\mu
  \left( \tau + \bar \tau \right) \left[ \dual{F}_{\mu\nu} + \tfrac{1}{2} \left( a - \bar a \right)  \hat\omega_{\mu\nu} \right] = 0 \, ,
\end{multline}
where \( \dual{F} = *_4 F \).

\paragraph{The scalar equation.}

Next we turn our attention to evaluating the scalar equation.   The main new ingredient with respect to the calculation in the literature ~\cite{Lambert:1997dm} is the presence of a \emph{rhs} term in Equation~\eqref{eq:general-scalar-equation}, which reads
\begin{equation}
  -\frac{2}{3 } \hat G\indices{^{I}_{mnp}} h^{mnp}
  = \frac{2}{ 1 + \abs{\del s}^2}\hat\omega^-_{\mu\nu} \cF_{\mu\nu} \left( \frac{\di a}{\di u}  \right)^{-1} \lambda_z + \frac{2}{ 1 + \abs{\del s}^2} \hat\omega^+_{\mu\nu} \wt{\cF}_{\mu\nu} \left( \frac{\di \bar a}{\di \bar u}  \right)^{-1} \bar \lambda_{\bar z} \ ,
\end{equation}
for both non-trivial cases \( X^{I} = s \) and \( X^{I} = \bar s \). The two corresponding scalar equations take the form
\begin{align}
  E &
  \begin{multlined}[t][.8\textwidth]
    =\del_\mu \del_\mu s - \del \left[ \frac{\del_\rho s\del_\rho s \bdel \bar s }{1 + \abs{\del s}^2} \right] - \frac{16 \del^2 s}{\left( 1 + \abs{\del s}^2 \right)^2}h_{\mu\nu \bar z} h_{\mu\nu \bar z}  \\
    -2\hat\omega^-_{\mu\nu} \cF_{\mu\nu} \left( \frac{\di a}{\di u}
    \right)^{-1} \lambda_z + 2 \hat\omega^+_{\mu\nu}
    \wt{\cF}_{\mu\nu} \left( \frac{\di \bar a}{\di \bar u}
    \right)^{-1} \bar \lambda_{\bar z} = 0 \, ,
  \end{multlined}\\
  \bar E &
  \begin{multlined}[t][.8\textwidth]
    = \del_\mu \del_\mu \bar s - \bdel \left[ \frac{\del_\rho \bar s \del_\rho \bar s \del s }{1 + \abs{\del s}^2} \right] - \frac{16 \bdel^2 \bar s}{\left( 1 + \abs{\del s}^2 \right)^2}h_{\mu\nu z} h_{\mu\nu z}  \\
    -2\hat\omega^-_{\mu\nu} \cF_{\mu\nu} \left( \frac{\di a}{\di u}
    \right)^{-1} \lambda_z + 2 \hat\omega^+_{\mu\nu}
    \wt{\cF}_{\mu\nu} \left( \frac{\di \bar a}{\di \bar u}
    \right)^{-1} \bar \lambda_{\bar z} = 0\;.
  \end{multlined}
\end{align}
In this case it is natural to integrate over the Riemann surface using the form \( \di z \wedge \bar \lambda \) and obtain the four-dimensional scalar equations of motion as
\begin{equation}
  \int_\Sigma E \di z \wedge \bar \lambda = \int_\Sigma \bar E \di \bar z \wedge \lambda = 0 \, .
\end{equation}
The details of the calculation are similar to those of the vector equation. The end result is 
\begin{align}
  \begin{multlined}[.9\textwidth]
    \left( \tau - \bar \tau \right) \del_\mu \del_\mu a + \del_\mu a \del_\mu \tau +  \frac{\di \bar \tau}{\di \bar a} \wt{\cF}_{\mu\nu} \wt{\cF}_{\mu\nu} \\
    -2 \left( \tau - \bar \tau \right) \hat\omega_{\mu\nu} \cF_{\mu\nu} + 2 \left( L_1 - L_2 \right) \left( \frac{\di \bar a}{\di \bar u} \right)^2 \hat\omega_{\mu\nu} \wt{\cF}_{\mu\nu} = 0 \; ,
  \end{multlined}\\
 \begin{multlined}[.9\textwidth]
   \left( \tau - \bar \tau \right) \del_\mu \del_\mu \bar a - \del_\mu \bar a \del_\mu \bar \tau - \frac{\di \tau}{\di a} \cF_{\mu\nu} \cF_{\mu\nu} \\
- 2 \left( \tau - \bar \tau \right) \hat\omega_{\mu\nu} \wt{\cF}_{\mu\nu} + 2 \left( \bar L_1 - \bar L_2 \right) \left( \frac{\di a}{\di u} \right)^2 \hat\omega_{\mu\nu} \cF_{\mu\nu } = 0\; ,
  \end{multlined}
\end{align}
where \( L_1  \) and \( L_2 \) are the integrals
\begin{align}
  L_1 &= -\int_\Sigma \del\left(\frac{\del s}{1 + \abs{\del s}^2}\right) (\bar s+\bar s - z\bdel\bar s - \bar z\bdel \bar s) \lambda_{\bar z} \di z \wedge   \bar \lambda \, ,\\
  L_2 &= \int_\Sigma \bar \lambda_{\bar z}  \di z \wedge   \bar \lambda \, .
\end{align}
The second integral can be evaluated straightforwardly  in terms of \( u \)  using the methods of Appendix~\ref{sec:line-integrals}:
\begin{equation}
  L_2 = \int_\Sigma \bar \lambda_{\bar z}^2 \di z \wedge \di \bar z  = {\pi \im}
  \left( \frac{ u - 1}{\abs{u - 1}} - \frac{  u + 1}{\abs{u + 1}} \right) \, .
\end{equation}
The evaluation of \( L_1 \) is more involved but leads to $L_1=L_2$ (see the appendix).

The scalar equations take the final form
\begin{align}
  \begin{multlined}[.9\textwidth]
    \label{eq:scalar-equation-a}
    \left( \tau - \bar \tau \right) \del_\mu \del_\mu a + \del_\mu a  \del_\mu \tau + 2 \frac{\di \bar \tau}{\di \bar a} \left( F_{\mu\nu} F_{\mu\nu} + F_{\mu\nu} \dual{F}_{\mu\nu} \right) \\
    + 4 \frac{\di \bar \tau}{\di \bar a} \left( a - \bar a \right) \hat\omega^+_{\mu\nu} F_{\mu \nu} - 4 \left( \tau - \bar \tau \right) \hat\omega^-_{\mu\nu} F_{\mu\nu} = 0\; ,
  \end{multlined}\\
  \begin{multlined}[.9\textwidth]
    \label{eq:scalar-equation-abar}
    \left( \tau - \bar \tau \right) \del_\mu \del_\mu \bar a - \del_\mu \bar a \del_\mu \bar \tau - 2 \frac{\di \tau}{\di a} \left( F_{\mu\nu} F_{\mu\nu} - F_{\mu\nu} \dual{F}_{\mu\nu} \right) \\
    + 4 \frac{\di \tau}{\di a} \left( a - \bar a \right) \hat\omega^-_{\mu\nu} F_{\mu \nu} - 4 \left( \tau - \bar \tau \right) \hat\omega^+_{\mu\nu} F_{\mu\nu} = 0\; .
  \end{multlined}
\end{align}

\paragraph{The four-dimensional action.}

It is well known that the equations of motion for a generic \M5 embedding do not stem from a six-dimensional action. On the other hand our calculation results in the four-dimensional equations of motion for the \( \Omega \)--deformation of the \ac{sw} theory, which we expect to have a Lagrangian description. In fact, a direct calculation shows that the vector equation~\eqref{eq:vector-equation} and the two scalar equations~\eqref{eq:scalar-equation-a} and~\eqref{eq:scalar-equation-abar} are all derived from the variation of the following Lagrangian:
\begin{multline}
   \im \mathscr{L} =- \left( \tau - \bar \tau \right) \left[ \tfrac{1}{2}\del_\mu a \del_\mu \bar a + F_{\mu\nu} F_{\mu\nu} + \left( a - \bar a \right) \dual{\hat\omega}_{\mu\nu} F_{\mu\nu} - 2 \del_\mu \left( a + \bar a \right) \hat\omega_{\mu \nu} A_\nu \right]  \\
  + \left( \tau + \bar \tau \right) \left[ F_{\mu\nu} \dual{F}_{\mu\nu} + \left( a - \bar a \right) \hat\omega_{\mu\nu} F_{\mu\nu} + 2 \del_\mu \left( a - \bar a \right) \hat\omega_{\mu \nu} A_\nu \right] \, .
\end{multline}
This is the main result of this paper and represents the \( \Omega \)--deformation of the \ac{sw} action. In this form the action is not manifestly gauge invariant. An equivalent, gauge invariant, form is given by
\begin{multline}
   \im  \mathscr{L} = - \left( \tau - \bar \tau \right)  \left[ \tfrac{1}{2} \del_\mu a \del_\mu \bar a + F_{\mu\nu} F_{\mu\nu} +  \left( a - \bar a \right) \dual{\hat\omega}_{\mu\nu} F_{\mu\nu} - 2 \del_\mu \left( a + \bar a \right) \dual{F}_{\mu \nu}  \dual{\hat U}_\nu \right] \\
   +  \left( \tau + \bar \tau \right)  \left[ F_{\mu\nu} \dual{F}_{\mu\nu} + \left( a - \bar a \right) \hat\omega_{\mu\nu} F_{\mu\nu} + 2 \del_\mu \left( a - \bar a \right) \dual{F}_{\mu \nu}  \dual{\hat U}_\nu \right] \, ,
\end{multline}
where $\omega = \di U$ and $\dual{\omega} = \di \dual{U}$.

Let us consider some generalizations of our calculation.  It is natural to write the action in a more supersymmetric form as a sum of squares:
\begin{multline}
 \im \mathscr{L} = - \left( \tau - \bar \tau \right) \Big[ \tfrac{1}{2}\left( \del_\mu a + \frac{2 \bar \tau }{\tau - \bar \tau} \dual{F}_{\mu\nu} \dual{\hat U}_\nu \right) \left( \del_\mu \bar a - \frac{2 \tau }{\tau - \bar \tau} \dual{F}_{\mu\nu} \dual{\hat U}_\nu \right) \\
  + \left( F_{\mu\nu} + \tfrac{1}{2} \left( a - \bar a \right) \dual{\hat\omega}_{\mu\nu} \right) \left( F_{\mu\nu} + \tfrac{1}{2} \left( a - \bar a \right) \dual{\hat\omega}_{\mu\nu} \right) \Big] \\
  +  \left( \tau + \bar \tau \right) \left( F_{\mu\nu} + \tfrac{1}{2} \left( a - \bar a \right) \dual{\hat\omega}_{\mu\nu} \right) \left( \dual{F}_{\mu\nu} + \tfrac{1}{2} \left( a - \bar a \right) \hat\omega_{\mu\nu}\right) \; .
\end{multline}
This therefore leads to a prediction for the ${\mathcal O}(\epsilon^2)$ terms. Note however that there could also be additional ${\mathcal O}(\epsilon^2)$ terms which are complete squares on their own, similar to the last term in (\ref{action}).

Finally, although our calculations were only performed in the simplest case of an $SU(2)$ gauge group with one modulus, it is natural to propose that the generalization to arbitrary gauge group and matter content is given by
\begin{multline}
 \im  \mathscr{L} = - \left( \tau_{ij} - \bar \tau_{ij} \right) \Big[ \tfrac{1}{2}\left( \del_\mu a^i + 2\left(\tfrac{  \bar \tau }{\tau - \bar \tau}\right)_{ik} \dual{F}^{k}_{\mu\nu} \dual{\hat U}_\nu \right) \left( \del_\mu \bar a^j -2 \left(\tfrac{  \tau }{\tau - \bar \tau}\right)_{jl} \dual{F}^{l}_{\mu\nu} \dual{\hat U}_\nu \right) \\
  + \left( F^i_{\mu\nu} + \tfrac{1}{2} \left( a^i - \bar a^i \right) \dual{\hat\omega}_{\mu\nu} \right) \left( F^j_{\mu\nu} + \tfrac{1}{2} \left( a^j - \bar a^j \right) \dual{\hat\omega}_{\mu\nu} \right) \Big] \\
+ \left( \tau_{ij} + \bar \tau_{ij} \right) \left( F^i_{\mu\nu} + \tfrac{1}{2} \left( a^i - \bar a^i \right) \dual{\hat\omega}_{\mu\nu} \right) \left( \dual{F}^{j}_{\mu\nu} + \tfrac{1}{2} \left( a^j - \bar a^j \right) \hat\omega_{\mu\nu}  \right)\ ,
\end{multline}
where we have used a suitable form for the inverse of $(\tau - \bar\tau)_{ij}$ which is taken to act from the left.

\section{Conclusions}\label{sec:conclusions}

In this paper we have computed the corrections to first order in $\epsilon$ to an \M5--brane wrapping a Riemann surface in the $\Omega$--background of~\cite{Hellerman:2011mv,Reffert:2011dp,Hellerman:2012zf}. The result can be viewed as the leading correction to the Seiberg--Witten effective action of $\mathcal{N}=2$ super-Yang--Mills theory with an $\Omega$--deformation.

The corrected effective action includes a shift in the gauge field strength as well as a sort of generalized covariant derivative for the scalar, including a non-minimal coupling to the gauge field.  A similar generalized covariant derivative  already appears in (\ref{action}) and is reminiscent of the equivariant differential used in~\cite{Nekrasov:2002qd}.

It is important to ask why the result we obtain, calculated as the classical motion of a single \M5--brane in M--theory, can capture quantum effects in four-dimensional gauge theory. To answer this we should restore the factors of $R$ and $\Lambda$ into the Riemann surface. This can be achieved by simply rescaling $\del s\to \Lambda^2 R\del s$, $\del s / \del a\to R  \del s / \del a$ and $\del s / \del u\to R  \del s / \del u$ along with their complex conjugates. However this replacement does not affect the final equations. On the other hand $R=g_sl_s$ can be related to the gauge coupling constant $g_4$ in the string theory picture. Thus the classical M--theory calculation in fact captures all orders of the four-dimensional gauge theory.

\section*{Acknowledgments}

We would like to thank Paul Howe for helpful correspondence.

\appendix

\section{Appendix: Non-holomorphic integrals over \( \Sigma \)}
\label{sec:line-integrals}

Most of the integrals over the Riemann surface \( \Sigma \) that appear in this note can be evaluated using the same strategy that consists in reducing them to line integrals, as in~\cite{Lambert:1997he}. As an example consider one of the integrals appearing in the vector equation:
\begin{equation}
  I = \int_\Sigma \del
  \left[ \frac{\del_\mu \bar s \del s}{1 + \abs{\del s}^2} \bar z \bdel \bar s   \right] \di z \wedge \bar \lambda \,.
\end{equation}
First we observe that \( \bar \lambda \) is an anti-holomorphic one-form, so we can write
\begin{equation}
  I = \int_\Sigma \di
  \left[ \frac{\del_\mu \bar s \del s}{1 + \abs{\del s}^2} \bar z \bdel \bar s    \right] \wedge \bar \lambda \,.
\end{equation}
From the explicit expression of \( s(z) \) one finds that the integrand has singularities  at the roots \( \bar e_i \) of \( Q(\bar z) \):
\begin{align}
  \bar e_i &= \pm \sqrt{\bar u\pm 1  } \ , & i &= 1, \dots, 4 \,.
\end{align}
For this reason we introduce a new surface \( \Sigma_\delta \) by cutting holes of radius \( \delta \) in \( \Sigma \) around \( e_i \). Then \( I \) becomes an integral over the boundary \( \del \Sigma_\delta \):
\begin{equation}
  I =   \oint_{\del \Sigma_\delta} \frac{\del_\mu \bar s \bdel \bar s}{1 + \abs{\del s}^2} \bar z \del s  \bar \lambda_{\bar z} \di \bar z \,.
\end{equation}
Since we are interested in the behaviour around \( e_i \) we can expand the integrand in powers of \( \delta \). Note that for \( z = e_i + \delta \),
\begin{equation}
  \frac{\abs{\del s}^2}{1 + \abs{\del s}^2} = \frac{1}{1 + 1/\abs{\del s}^2} = \frac{1}{1 + \abs{Q}/(4 \abs{z}^2)} = 1  + \mathcal{O}(\delta)  \,.
\end{equation}
Moreover, since \( \bar s (\bar z) \) depends on \( x^\mu \) only via the modulus \( \bar u \) (Eq.~(\ref{eq:xmu-dependence})), \(\del_\mu \bar s = \del_\mu \bar u \bar \lambda_{\bar z} \), and the integral takes the form
\begin{equation}
  I =   \del_\nu \bar u\sum_i \oint_{\gamma_i} \bar e_i \bar \lambda_{\bar z}^2 \di \bar z + \mathcal{O} (\delta) \ ,
\end{equation}
where \( \gamma_i \) is a circle of radius \( \delta \) around \( e_i \), and \( \del \Sigma_\delta = \cup_i \gamma_i \). From the explicit expression of \( \bar s \)  we find that
\begin{equation}
  \bar \lambda_{\bar z}^2 = \frac{1}{\bar Q (\bar z)} \, ,
\end{equation}
so that each integral around \( \gamma_i \) can be evaluated using the residue theorem:
\begin{equation}
  \oint_{\gamma_i} \frac{1}{\bar Q (\bar z)} \di \bar z = -\frac{2 \pi \im }{\prod_{j \neq i}  (\bar e_i - \bar e_j)} \ ,
\end{equation}
and the whole integral is given by
\begin{equation}
  I = -2 \pi \im   \del_\mu \bar u \sum_{i =1}^4 \frac{\bar e_i}{\prod_{j \neq i} (\bar e_i - \bar e_j)} \,.
\end{equation}
By using the explicit values of \( e_i  \) we finally find that \( I \) vanishes.

\bigskip 

\begin{figure}
  \centering
  \includegraphics[width=.7\textwidth]{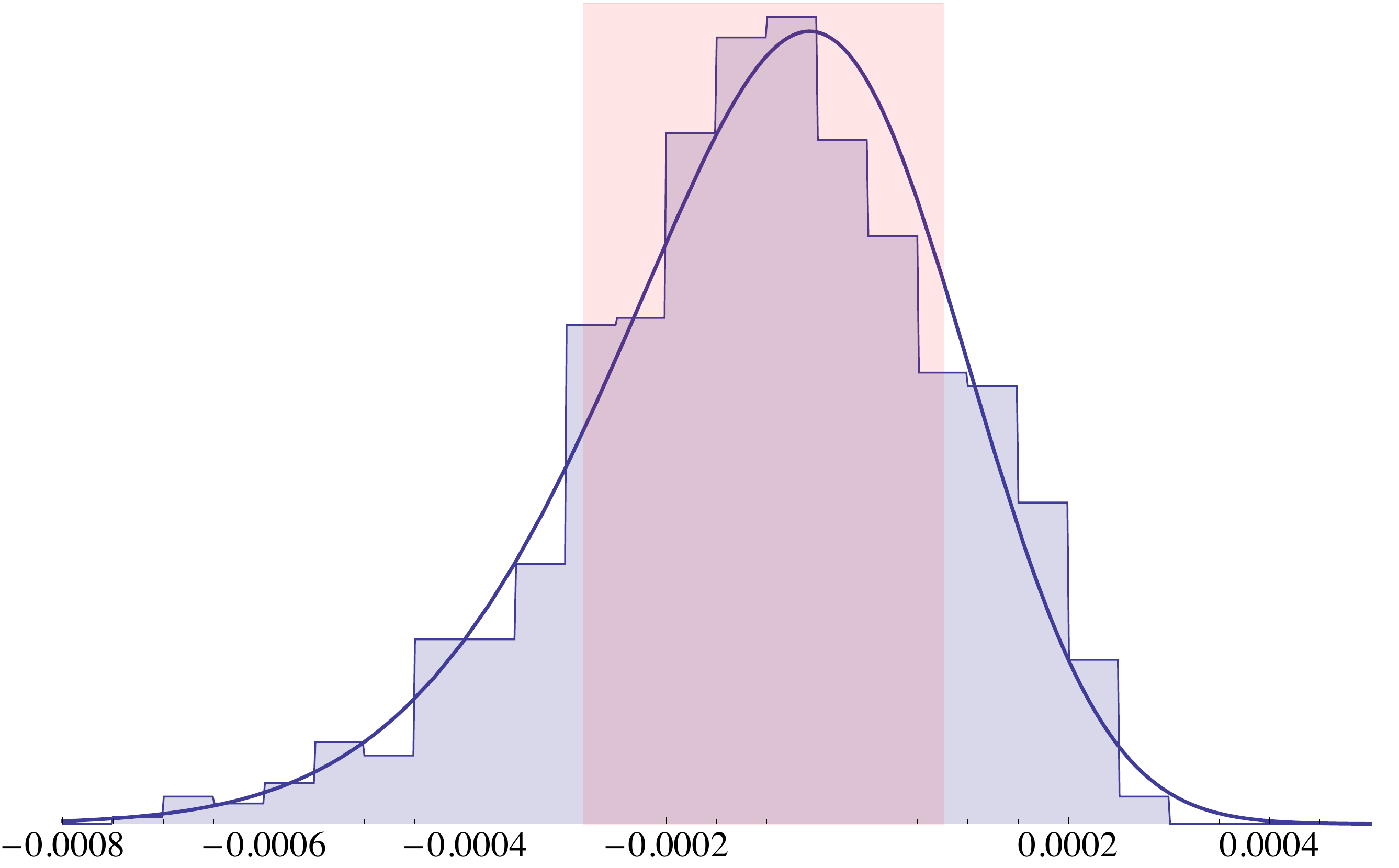}
  \caption{Numerical integration of \( L_2 \). The histogram collects the frequency the values of \( 1 - \abs{{L_1}/{L_2} } \) obtained by integrating for \( 10^3 \)  random values of \( u \). The continuous line is a skew normal distribution with average  \( -1.0 \times 10^{-4} \pm 1.7 \times 10^{-4} \) (pink region). The result is consistent with \( L_1 = L_2 \). We have also performed similar three-dimensional plots for the complex function $1-L_1/L_2$ which shows a clear peak around zero. }
  \label{fig:L2-Integral}
\end{figure}

Let us now examine the $L_1$ integral that appeared in the scalar equation. First we integrate by parts:
\begin{equation}
  \label{evalL1}
  \begin{aligned}
    L_1 ={}& -\int_\Sigma \di\left(\frac{\del s}{1 + \abs{\del s}^2}\right) ( s+\bar s - z\bdel\bar s - \bar z\bdel \bar s) \lambda_{\bar z}   \wedge   \bar \lambda \\
    ={}& -\oint_{\del\Sigma_\delta}\frac{\del s(\bar s+\bar s - z\bdel\bar s - \bar z\bdel \bar s) }{1 + \abs{\del s}^2} \lambda_{\bar z}^2 \di\bar z  +\int_\Sigma \frac{(\del s)^2 - \abs{\del s}^2}{1 + \abs{\del
        s}^2} \lambda_{\bar z}^2 \di z \wedge \di\bar z\ .
  \end{aligned}
\end{equation}
Using similar techniques to the $I$ integral above one finds that the boundary term is
\begin{equation}
  \begin{aligned}
    -\oint_{\del\Sigma_\delta}\frac{\del s(\bar s+\bar s - z\bdel\bar s - \bar z\bdel \bar s) }{1 + \abs{\del s}^2} \lambda_{\bar z}^2 \di\bar z &=   -{ 2 \pi \im }\sum_{i =1}^4 \frac{e_i}{\prod_{j \neq i}  (  \bar e_i - \bar e_j)} \\
    &= {\pi \im} \left( \frac{ u - 1}{\abs{u - 1}} - \frac{  u + 1}{\abs{u + 1}} \right)\\ &= L_2 \ .
  \end{aligned}
\end{equation}
Let us now look at the last term of~\ref{evalL1}. Rewriting   the integrand in terms of $Q$ we find
\begin{equation}
  \label{evalL1a}
  \begin{aligned}
    \int_\Sigma  \frac{(\del s)^2 - \abs{\del s}^2}{1 + \abs{\del s}^2} \lambda_{\bar z}^2 \di z \wedge  \di\bar z &= \int_\Sigma \frac{\abs{z}^2}{\frac{1}{4}\abs{Q}+\abs{z}^2}\left(\frac{z}{\bar z}- \sqrt{\frac{Q}{\bar Q}}\right) \frac{\di z}{\sqrt{Q}} \wedge  \frac{\di\bar z}{\sqrt{\bar Q}} \\
    &= \frac{1}{4}\int_\Sigma \frac{1}{1+\abs{z'/z}^2} \frac{z}{\bar
      z}\di y \wedge \di\bar y - \frac{1}{4}\int_\Sigma
    \frac{1}{1+\abs{z/z'}^2}{\frac{z'}{\bar z'}} \di y \wedge \di\bar
    y \ ,
  \end{aligned}
\end{equation}
where we changed variables to $\di y = 2 \di z/\sqrt{Q}$ so that  $z$ is now a holomorphic function of $y$ with $z' = \di z/ \di y $. We will now show that both terms on the \emph{rhs} vanish separately. Consider the first term  on the \emph{rhs} and expand  in a power series of $\abs{z'/z}$:
\begin{equation}
  \int_\Sigma \frac{1}{1+\abs{z'/z}^2} \frac{z}{\bar z}     \di y \wedge   \di\bar y = \sum_{n=0}^\infty \int_\Sigma (-1)^n \abs{\frac{z'}{z }}^{2n}  \frac{z}{\bar z} \di y \wedge   \di\bar y\ .
\end{equation}
Unfortunately  the \emph{rhs} here is not well-defined, even though the \emph{lhs} is. To correct this we can introduce two-step regulator with parameters $a$ and $b$ which we will later set to zero.  Thus we instead consider
\begin{equation}
  \begin{aligned}
    \MoveEqLeft[5] \int_\Sigma \frac{\eu^{- \abs{z'/z}^2a^2}\eu^{-b^2(\abs{z}^2 +1/\abs{z}^2) }}{1+\abs{z'/z}^2}  \frac{z}{\bar z}  \di y \wedge   \di\bar y\\
    ={} & \sum_{n=0}^\infty \int_\Sigma (-1)^n\abs{\frac{z'}{z }}^{2n}\eu^{- \abs{z'/z}^2a^2}\eu^{-b^2(\abs{z}^2 +1/\abs{z}^2) } \frac{z}{\bar z}  \di y \wedge   \di\bar y \\
    ={} & \sum_{n=0}^\infty (-1)^n\int_\Sigma  \frac{z}{\bar z}  \eu^{- \abs{z'/z}^2a^2}\eu^{-b^2(\abs{z}^2 +1/\abs{z}^2) } \di y_n \wedge   \di\bar y_n\ ,
\end{aligned}
\end{equation}
where we have changed variables  again to  $\di y_n = (z'/z )^n \di y$. Let us now set $a=0$ to deduce that
\begin{equation}
 \int_\Sigma \frac{ \eu^{-b^2(\abs{z}^2 +1/\abs{z}^2) }}{1+\abs{z'/z}^2}  \frac{z}{\bar z}    \di y \wedge   \di\bar y   = \sum_{n=0}^\infty (-1)^n\int_\Sigma  \frac{z}{\bar z}   \eu^{-b^2(\abs{z}^2 +1/\abs{z}^2) } \di y_n \wedge   \di\bar y_n\ .
\end{equation}
  In each of the terms of the sum $z$ is a holomorphic function of $y_n$ and
therefore $z(y_n)$ covers the whole complex plane (with the exception of one point) and
hence the integral of the phases $z/\bar z$  must vanish since the $b$-regulator is independent of the  phase. We can now set $b=0$ to see that each term in the sum vanishes and hence the first term  on the \emph{rhs} of ~\ref{evalL1a} vanishes. Finally we can repeat a similar argument for the second term on the \emph{rhs} of~\ref{evalL1a} only in this case the $b$-regulator should be taken to be $\eu^{-b^2(\abs{z'}^2 +1/\abs{z'}^2) }$. Thus we see that~\ref{evalL1a} vanishes and hence $L_1=L_2$.
The above proof that $L_1=L_2$ is   a little suspect since we required two regulators and needed to set $a=0$ first and then $b=0$. As a check we performed a numerical integration for random values of  \( u \) which clearly supports our claim (see Figure~\ref{fig:L2-Integral}).

\printbibliography

\end{document}